\documentclass[aps,prb,showpacs,floatfix,floats,superscriptaddress,twocolumn]{revtex4}
\usepackage{times}
\usepackage{bm}
\usepackage{subfig}
\usepackage{graphicx}

\newcommand{\be}{\begin{equation}}
\newcommand{\ee}{  \end{equation}}
\newcommand{\ba}{\begin{eqnarray}}
\newcommand{\ea}{  \end{eqnarray}}
\newcommand{\bi}{\begin{itemize}}
\newcommand{\ei}{  \end{itemize}}

\begin{document}

\title{Conductivity and Fano factor in disordered graphene}

\author{C. H. Lewenkopf}
\affiliation{Departamento de F\'{\i}sica Te\'orica, Universidade do Estado 
do Rio de Janeiro,
20550-900 Rio de Janeiro, Brazil}
\affiliation{Harvard University, Department of Physics, Cambridge, 
Massachusetts 02138, USA}
\author{E. R. Mucciolo}
\affiliation{Department of Physics, University of Central Florida, P.O. Box 
162385, Orlando, Florida 32816, USA}
\author{A. H. Castro Neto}
\affiliation{Department of Physics, Boston University, 590 Commonwealth 
Avenue, Boston, Massachusetts 02215, USA}

\date{\today}

\begin{abstract}
Using the recursive Green's function method, we study the problem of
electron transport in a disordered single-layer graphene sheet. The
conductivity is of order $e^2/h$ and its dependence on the carrier
density has a scaling form that is controlled solely by the disorder
strength and the ratio between the sample size and the correlation
length of the disorder potential. The shot noise Fano factor is shown
to be nearly density independent for sufficiently strong disorder,
with a narrow structure appearing at the neutrality point only for
weakly disordered samples. Our results are in good agreement with
experiments and provide a way for extracting microscopic information
about the magnitude of extrinsic disorder in graphene.
\end{abstract}


\maketitle

Graphene, a two-dimensional (2D) allotrope of carbon with a honeycomb
lattice, has attracted a lot of attention due to its unusual,
Dirac-like, electronic spectrum and its potential for an all-carbon
based electronics.\cite{geim_review} While the theoretical literature
on graphene is already quite extensive,\cite{RMPCastroNeto} the effect
of disorder on transport properties near the charge neutrality point
(Dirac point) is still subject to much debate and controversy. The
difficulty in understanding how disorder affects Dirac fermions has to
do with the fact that electrochemical disorder is a relevant
perturbation under the renormalization group and drives the system
away from the weak disorder regime.\cite{Ludwig94} Hence, standard
perturbation theory fails and one has to rely on either
non-perturbative methods or numerical approaches. Although several
recent theoretical works have been established that for short-range
disorder large graphene samples should become
insulators,\cite{Altland06} the current understanding of long-range
disorder is less clear. Some authors \cite{NomuraMacDonald07} have
questioned the existence of a beta function for undoped graphene while
others \cite{Ostrovsky07} have proposed a non-monotonic beta function
for single-valley Dirac fermions and a metal-insulator
transition. Numerical computations in momentum space \cite{Nomura07}
as well as simulations based on a transfer-matrix method
\cite{Bardarson07,San-Jose07} adapted to the single-valley Dirac
equation \cite{Titov07} found instead a simple scaling law for the
conductivity, a metallic beta function, and no indication of a new
fixed point.\cite{ryu07} Furthermore, while the self-consistent Born
approximation (SCBA) predicts a universal (impurity independent)
conductivity of $4e^2/(\pi h)$,\cite{Fradkin} percolation theory finds
a smaller value.\cite{falko}

On the experimental side, graphene behaves as a good metal with
conductivities of the order of $e^2/h$,\cite{experiments} which is
inconsistent with a purely ballistic transport. No sign of strong
localization has been seen in graphene although the size of the
samples (a few micrometers) may be smaller than the localization
length in 2D.

Current experiments indicate that transport is not ballistic, but is
it really diffusive? Experimentally, the electron mean free path
$\ell_{\rm tr}$ can be estimated by gating graphene away from the
Dirac point and using the Drude formula to relate $\ell_{\rm tr}$ to
the conductivity $\sigma$: $\ell_{\rm tr} = \sigma/[(2e^2/h)\sqrt{\pi
n}]$, where $n$ is the carrier density. Typically, $\ell_{\rm tr}$ is
found to be smaller but close to the sheet linear size $L$. Although
this estimate has to be taken with caution, it supports the notion
that current experiments are in the crossover regime between ballistic
and diffusive, a regime where standard analytical approaches
fail. Establishing a quantitatively understanding of the effect of
disorder in the transport properties of graphene is of fundamental
importance for the development of electronic devices where high
mobilities are required. Therefore, an alternative theoretical study
is desirable.

\begin{figure}[tp]
\centering
\subfloat{
\label{fig:pot}
\includegraphics[width=1.75in]{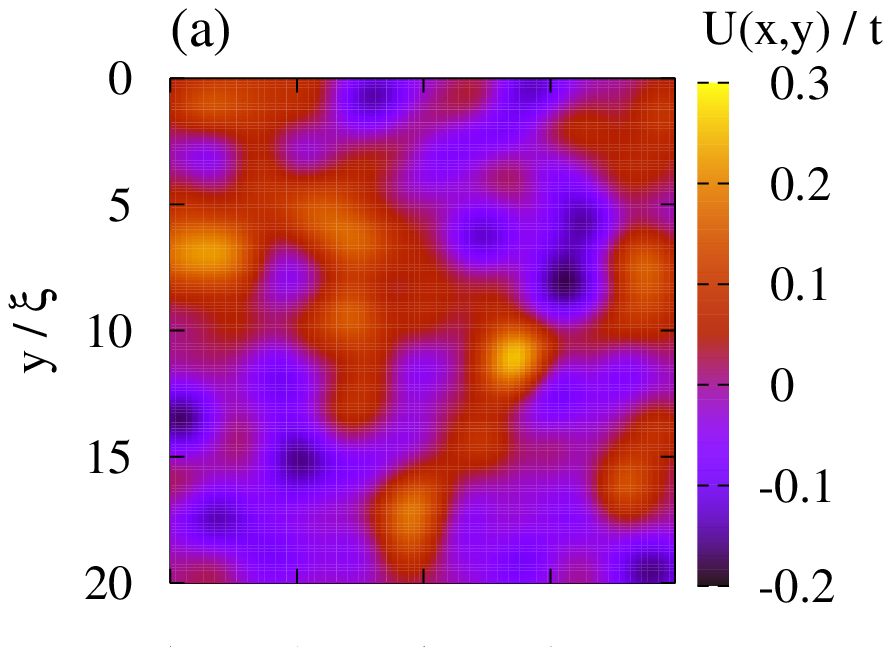}}\hspace{-.25in}
\subfloat{
\label{fig:LDOSb}
\includegraphics[width=1.75in]{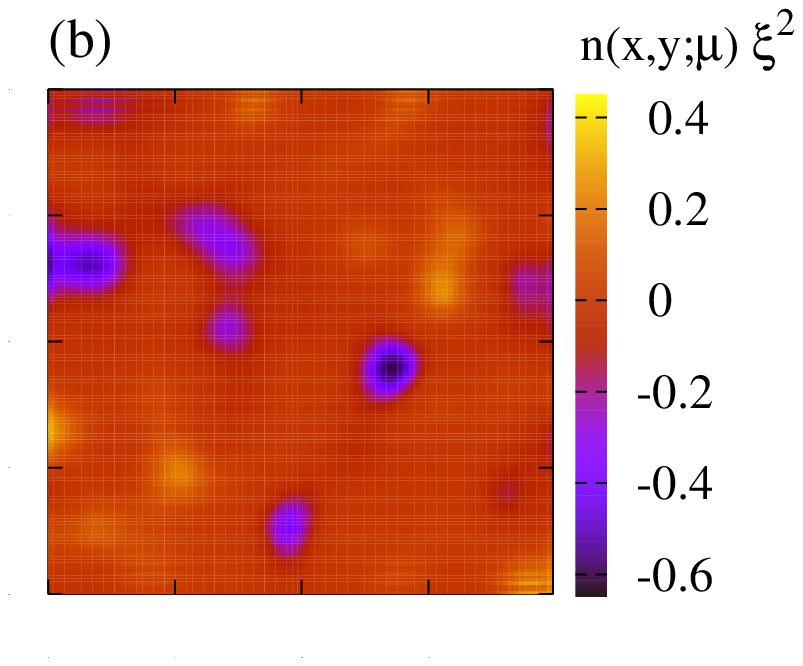}}
\\
\vspace{-.3in}
\subfloat{
\label{fig:LDOSc}
\includegraphics[width=1.75in]{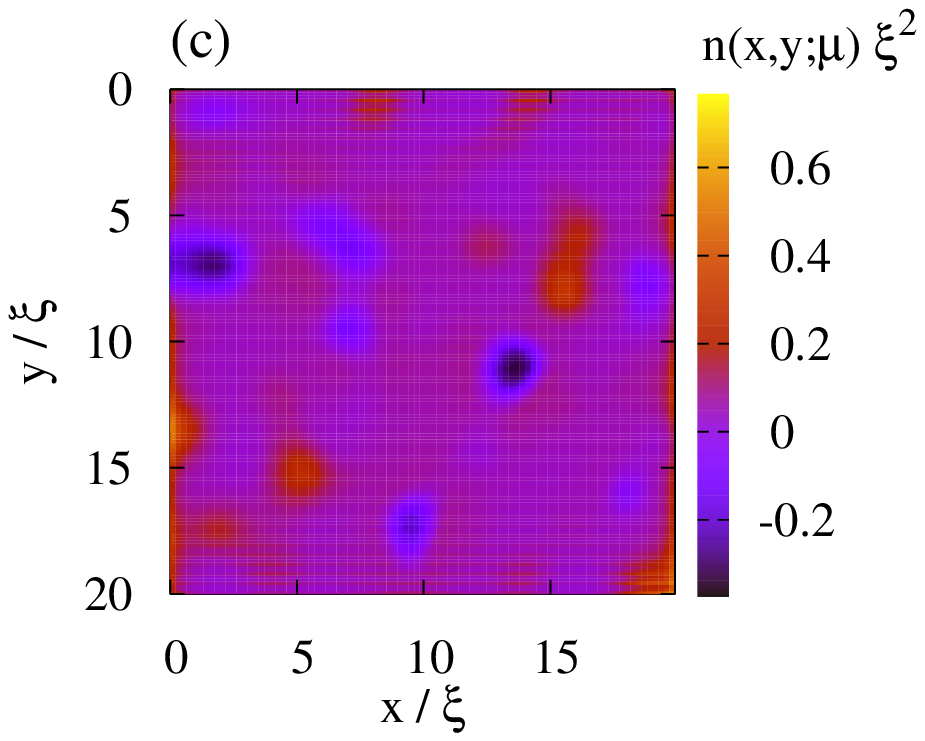}}\hspace{-.25in}
\subfloat{
\label{fig:LDOSd}
\includegraphics[width=1.75in]{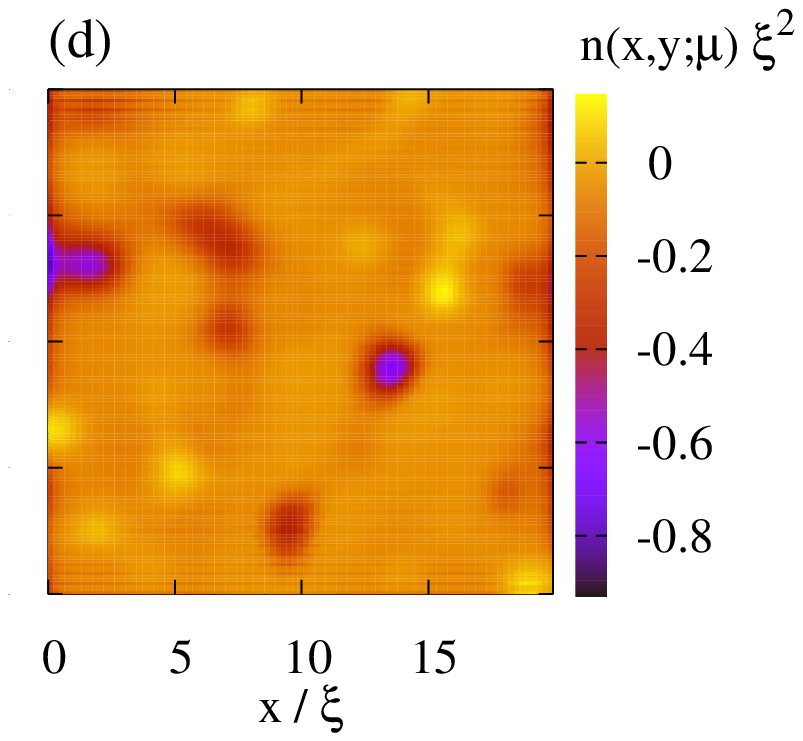}}
\captionsetup{justification=RaggedRight,singlelinecheck=false}
\caption[]{(Color online) Color maps: \subref{fig:pot} Long-range
disordered potential in a square graphene sheet ($M=161$, $N=92$,
$K_0=1$, $n_{\rm imp} = 0.02$, $\xi=4a_0$); Carrier densities
(positive for electron and negative for holes) at the Dirac point
\subref{fig:LDOSb}, above it \subref{fig:LDOSc}, and below
\subref{fig:LDOSd}, with $\mu/t = 0$, $0.1$, and $-0.1$,
respectively.}
\label{fig:LDOS}
\end{figure}

In this paper, we use the recursive Green's function method
\cite{LeeFisher81} to study numerically the effect of extrinsic
long-range disorder in the linear electronic transport of graphene at
low carrier densities. We explore the low temperature limit, when
intrinsic scattering mechanisms such as electron-phonon can be
neglected. Our calculations are based on a microscopic, tight-binding
model of the graphene sheet and performed at and near the Dirac
point. After a simple rescaling, we find that the dependence of the
conductivity on the carrier density is controlled only by a
dimensionless disorder amplitude and the ratio of system size to
disorder correlation length, $L/\xi$. This dependence can vary from
sublinear to superlinear, depending upon the disorder
strength. Therefore, our results provide a way for extracting
extrinsic disorder parameters from experimental data without making
any assumption about the underlying dynamical regime. We also find
that disorder has a marked effect on how the shot noise Fano factor
depends on the carrier density. For samples with square aspect ratios
and weak disorder, a narrow dip exists at the Dirac point, while for
sufficiently strong disorder, the Fano factor becomes density
independent, taking up values larger than 1/3 which increase slowly
with the disorder.

The nearest-neighbor tight-binding model on a honeycomb lattice with
site disorder reads
%
$
H = - \sum_{\langle i,j \rangle} t |i\rangle \langle j| + \sum_i
U({\bf r}_i)\, |i\rangle \langle i|,
$
%
where $i$ and ${\bf r}_i$ label the lattice sites and their
coordinates, respectively, and $t\approx 2.8$ eV is the hopping
energy.  The long-range disordered potential $U({\bf r}_i)$ consists
of ${\cal N}_{\rm imp}$ random lattice sites $\{{\bf R}_n\}$. Each of
these sites is the center of a Gaussian scatterer with a random
amplitude $U_n$ taken from a uniform distribution over the interval
$[-\delta,\delta]$:
%
$
U({\bf r}_i) = \sum_{n=1}^{{\cal N}_{\rm imp}} U_n\, e^{-|{\bf
r}_i - {\bf R}_n |^2/2 \xi^2},
$
%
where $\xi > a_0$ is the range of the potential ($a_0 \approx 2.46$
\AA\ is the lattice constant).\cite{ShonAndo98,Rycerz06} A typical
realization of the disorder is shown in
Fig. \ref{fig:LDOS}\subref*{fig:pot}. The concentration of scatterers
is $n_{\rm imp} = {\cal N}_{\rm imp}/{\cal N}$, where ${\cal N}$
denotes the total number of atomic sites. When $n_{\rm imp} \ll 1$,
the magnitude of the disorder fluctuations is characterized by the
dimensionless parameter $K_0$, which is defined from the impurity
potential correlation function
\begin{equation}
\label{eq:disorder}
\langle U({\bf r}_i)\, U({\bf r}_j) \rangle = \frac{K_0 (\hbar
v)^2}{2\pi \xi^2}\, e^{-|{\bf r}_i - {\bf r}_j|^2/2\xi^2},
\end{equation}
where $v = \sqrt{3}\, a_0\, t/2\hbar$ is the Fermi velocity (notice
that $\langle U({\bf r}_i)\rangle=0$). We note that $K_0$ contains
information not only about the relative magnitude of the potential
fluctuations, $\delta/t$, but also about the scatterers' range and
concentration: A simple calculation yields $K_0 \approx 40.5\, n_{\rm
imp} (\delta/t)^2 (\xi/a_0)^4$.\cite{Rycerz06} In the continuum limit
and using Eq. (\ref{eq:disorder}) together with the Born approximation
(BA), one finds that the transport mean free path away from the Dirac
point is given by $\ell_{\rm tr}^{\rm BA} = 2\lambda_F/(\pi K_0)$,
where $\lambda_F$ is the Fermi wavelength in the graphene sheet
\cite{ShonAndo98} ($\lambda_F \ll \ell_{\rm tr}^{\rm BA}$ for the BA
to hold).

We investigate square graphene sheets with $M = 1+2L/a_0$ and $N =
1+2\sqrt{3}L/a_0$ lattice sites along perpendicular edges. The sheet
is connected to ballistic, square lattice source and drain leads
following standard procedures.\cite{LeeFisher81,schomerus07} For all
data shown here, the edges parallel to the electron current flow are
in the metallic armchair configuration \cite{BreyFertig06} (we have
found that, in the presence of disorder, similar results are obtained
for the zigzag configuration). Using the recursive method, we compute
the electron retarded Green's function at any lattice site, $G^r({\bf
r}_i;E)$, where $E$ is the electron energy. The local density of
states, $\nu({\bf r}_i;E) = -(1/\pi) {\rm Im} G({\bf r}_i;E)$, is
obtained and can be integrated in energy to yield the local carrier
density at the Fermi energy, $n({\bf r}_i;\mu) = \int_{U({\bf
r}_i)}^\mu dE\, \nu({\bf r}_i;E)$. This is illustrated in
Figs. \ref{fig:LDOS}\subref*{fig:LDOSb}-\subref*{fig:LDOSd}, where we
show the local carrier density for a fixed realization of the disorder
potential and three values of the Fermi energy. The hole and electron
puddles formed at the maxima and minima of the potential are clearly
visible. These patterns are reminiscent of the single-electron
transistor scannings of graphene.\cite{yacoby}

\begin{figure}[t]
\includegraphics[width=0.95\columnwidth]{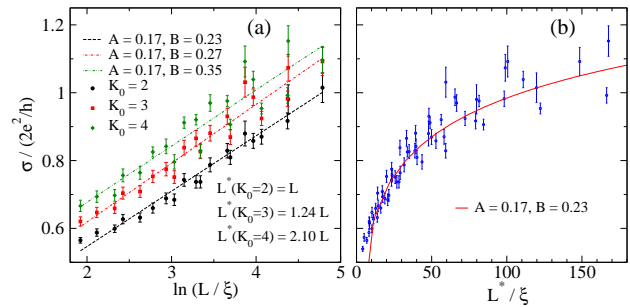}
\captionsetup{justification=RaggedRight,singlelinecheck=false}
\caption{(Color online) (a) Scaling of the average conductivity with
sample size for a square graphene sheet at the Dirac point. The data
corresponds to $M = 41, 83, 113, 161, 317, 479$, $\xi/a_0 = 2, 3, 5$,
and $n_{\rm imp} = 0.02$. The lines denotes the best-fits of the form
$\sigma/(2e^2/h) = A \ln(L/\xi) + B$ to the data. (b) Average
conductivity versus rescaled sheet lengths, $L^\ast = \alpha(K_0)\,
L$, with $\alpha$ determined from the best-fits found in (a).}
\label{fig:Dirac-conductance}
\end{figure}

The recursive Green's function method provides a efficient way to
compute the electron's scattering matrix as a function of the electron
energy, $S(E)$. From $S$, one obtains the transmission $t$ and
reflection $r$ matrices, which are then used in the computation of the
linear conductance ${\cal G}$ and the shot noise Fano factor $F$:
${\cal G} = (2e^2/h) {\rm Tr} (t t^\dagger)$ and $F = {\rm
Tr}(t^\dagger t \,r^\dagger r)/{\rm Tr}(t^\dagger t)$ (the factor of 2
in ${\cal G}$ accounts for spin degeneracy). We note that $\sigma =
{\cal G}$ for a square sheet.

In Fig.~\ref{fig:Dirac-conductance}, we show the scaling of the
conductivity with $L$ at the Dirac point for several disorder
strengths and correlation lengths. Our results are in qualitative
agreement with recent numerical studies using the continuum model
\cite{Nomura07,Bardarson07,San-Jose07} and are compatible with a
metallic beta function of the form $\beta(\sigma) = A\,
(2e^2/h)/\sigma$, with $A\approx 0.17$. In particular, the
quantitative agreement with the scaling obtained in
Refs. \onlinecite{Nomura07,Bardarson07} indicates that for correlation
lengths $\xi \agt 2a_0$, there is no noticeable effect of intervalley
scattering. Thus, the lattices in our study are large enough to
reproduce the numerical results obtained with the continuous
single-valley model.

In order to explore how the conductivity behaves as a function of
carrier density $n$, we compute the latter as a function of the Fermi
energy $\mu$ in the graphene sheet through the expression $n(\mu) =
\int_0^{\mu} dE \,\nu(E)$, where $\nu(E)$ is the sheet global density
of states (DOS), which can be readily obtained from the energy
dependence of the scattering matrix. By writing the scattering matrix
in terms of the retarded Green's functions,\cite{LeeFisher81} one can
show that the Wigner-Smith time delay, $\tau_{\rm W}(E) = - i
\hbar(d/dE)\ln( \det S)$, becomes $\tau_{\rm W}(E) = - 2\hbar\, {\rm
Tr \,[Im\,} G^r({\bf r}_i; E)]$. We thus arrive at $\nu(E) = - (i/\pi
WL) {\rm Tr} \left[ S^\dagger (dS/dE) \right]$.  These steps show that
$\nu = 2 \tau_{\rm W}/(hWL)$, as expected from scattering theory.

\begin{figure}[t]
\includegraphics[width=0.95\columnwidth]{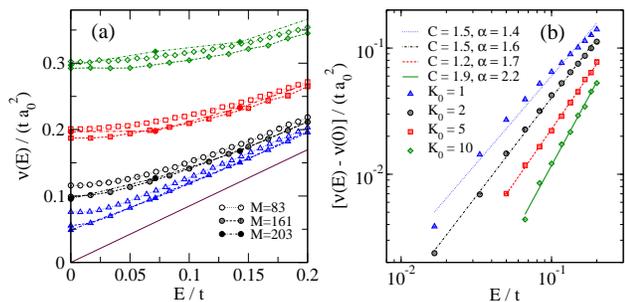}
\captionsetup{justification=RaggedRight,singlelinecheck=false}
\caption{(Color online) (a) Average density of states for three
lattice sizes (different symbols) and four values of $K_0$ (different
colors and line types). Notice that $L=(M-1)a_0/2$. Only data for
$E>0$ is shown. The thick solid line represents the case of a clean,
infinite graphene sheet and the other lines are guides to the
eyes. The number of realizations varies from 100 to 10,000 with $\xi =
2a_0$ and $n_{\rm imp} = 0.02$. (b) Shifted density of states for
$M=161$. The lines correspond to best-fits of the form $\nu(E) -
\nu(0) = C\, t\, a_0^2\, |E/t|^\alpha$.}
\label{fig:DOS-n-scaled}
\end{figure}

In Fig.~\ref{fig:DOS-n-scaled}a, we show the average $\nu(E)$ for
different values of the disorder parameter $K_0$ and a fixed
$\xi$. Notice that the DOS does not vanish at the Dirac point, as one
expects for an infinite clean graphene sheet. The coupling to the
semi-infinite leads broadens the electronic energy levels in the
graphene sheet, an effect that is more pronounced in the vicinity of
the Dirac point. More interesting however is the effect of
disorder. We observe that as $K_0$ increases, the DOS becomes almost
flat over a large energy range. In particular, close to the Dirac
point the DOS grows substantially with $K_0$, surpassing the
broadening due to the coupling to the leads.

Remarkably, we find numerical evidence that the DOS is very robust
against system size scaling. Figure \ref{fig:DOS-n-scaled}a shows that
$\nu(E)$ rapidly converges as $M$ is increased and $K_0<10$. Another
feature worth noticing is that, for strong enough disorder, $\nu(E)$
grows approximately as a power-law with $E$, as shown in
Fig.~\ref{fig:DOS-n-scaled}b. Our results for the DOS complement the
analysis of Ref. \onlinecite{Ludwig94} of the problem of Dirac
fermions in the presence of a random chemical potential. This kind of
disorder was shown to be a relevant perturbation in the
renormalization group sense, possibly taking the system to a new fixed
point if electron-electron interactions are taken into
account.\cite{Ostrovsky07}

\begin{figure}[t]
\includegraphics[width=0.85\columnwidth]{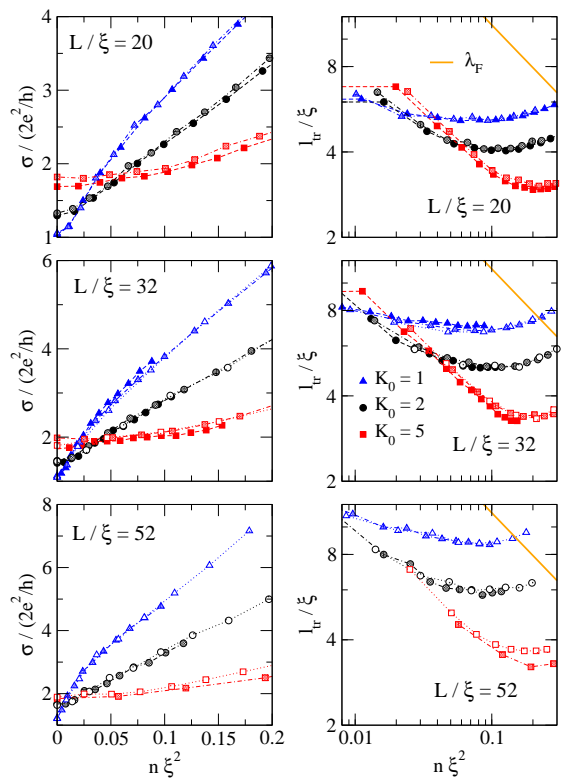}
\captionsetup{justification=RaggedRight,singlelinecheck=false}
\caption{(Color online) Average conductance (left column) and Drude
transport mean free path (right column) as functions of carrier
density for several disorder strengths ($K_0 = 1,2,5$), lattice sizes
($L/a_0 = 80,101,164$), and disorder correlation lengths ($\xi/a_0 =
2,2.5,3,3.3,4,5$). The number of realizations varies from 100 to 5,000
depending on $K_0$ and $L$. The thick solid line corresponds to the
clean system Fermi wavelength, $\lambda_F=2\sqrt{\pi/n}$. The
transport mean free path is calculated using $\ell_{\rm tr} =
\sigma/[(2e^2/h) \sqrt{\pi n}]$.}
\label{fig:VIInew}
\end{figure}

We now turn our attention to transport properties. Our main results
are collected in Fig. \ref{fig:VIInew}. As one departs from the Dirac
point, the counterintuitive feature that the conductivity is enhanced
by disorder disappears.\cite{Titov07} For a fixed disorder, the
conductivity increases with carrier density. By plotting $\sigma$ as a
function of the number of electrons contained in a square of area
$\xi^2$, we find that the functional dependence is solely controlled
by the disorder strength $K_0$ and the ratio $L/\xi$. Notice that
$\sigma$ as a function of $n\xi^2$ goes from sublinear (weak disorder)
to linear (intermediate disorder) and then to superlinear (strong
disorder) as $K_0$ is varied. All three behaviors are seen in the
experiments \cite{experiments} and therefore may be related to the
strength of the long-range disorder in the samples. For a fixed $K_0$
and dilute scatterers, we found little dependence of our results on
$n_{\rm imp}$.

The scaling behavior of $\sigma = \sigma(n\xi^2, K_0, L/\xi)$ works in
the crossover regime from ballistic to diffusive regimes and departs
strongly from the SCBA prediction,\cite{ShonAndo98} which is only
justifiable in the semiclassical diffusive regime, when $\lambda_F \ll
\ell_{\rm tr}^{\rm BA} \ll L$. Notice that $\ell_{\rm tr} \sim \alt
\lambda_F$ in Fig. \ref{fig:VIInew} and therefore our simulations are
far from the semiclassical limit. Although the lattices investigated
are rather small in comparison to the graphene samples probed in
experiments, the scaling allows us to make a direct comparison with
experimental data by using realistic values of $\xi$ (which ranges
between 10 and 100 nm, as shown in Ref. \onlinecite{yacoby}) and
extrapolating the $\sigma$ curves to large carrier densities,
typically of order $10^{12}$ cm$^{-2}$. This in turn provides a way
for estimating the value of $K_0$ in transport measurements dominated
by extrinsic disorder. As an example, in Fig. \ref{fig:VIInew}, if we
take $L/\xi=20$ and $K_0=2$ (which yields an approximately linear
$\sigma(n)$ dependence) and assume $\xi= 50$ nm, we obtain $\sigma
\approx 500\, e^2/h$, which is just a bit higher than the values
measured experimentally in high mobility samples.

\begin{figure}[t]
\includegraphics[width=0.9\columnwidth]{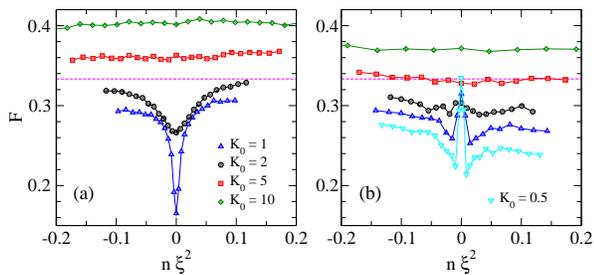}
\captionsetup{justification=RaggedRight,singlelinecheck=false}
\caption{(Color online) Average Fano factor as a function of carrier
density for several disorder strengths and $\xi = 2a_0$. The number of
realizations ranges from 100 to 5,000. (a) Square sheets ($W/L=1$)
with $L/\xi=40$. (b) Rectangular sheets ($W/L = 3.1$) with $L/\xi=26$.
The dashed lines represent $F=1/3$.}
\label{fig:FigFano}
\end{figure}

In Fig. \ref{fig:FigFano}, we present the typical dependence of the
average Fano factor on the carrier density and disorder strength for
square and rectangular sheets. For weakly disordered square samples,
we observe a narrow dip at the Dirac point followed by a saturation of
$F$ as $n$ is increased. As the disorder increases, the dip rapidly
disappears and $F$ takes a density-independent value above 1/3, which
is the value expected for a normal metal in the diffusive
regime.\cite{blanter} In contrast to the behavior seen for the
conductivity, we do not observe a clear scaling behavior for $F$,
namely, the data do not collapse onto distinct universal curves when
sorted out by $K_0$ and $L/\xi$ (not shown). For short and wide
rectangular sheets, the behavior is quite distinct from the square
case. For weak disorder, there is a peak located at the Dirac point,
followed by a damped oscillation towards a saturation value (the
asymmetry is due to a band mismatch between the honeycomb and lead
lattices). These features are compatible with analytical results
derived for clean junctions with large $W/L$ aspect
ratios.\cite{Tworzydlo06} As the disorder strength increases, the peak
widens, the oscillations are smeared, and the saturation value of $F$
goes up. For large disorder strengths, $F$ becomes again nearly
density independent. For both geomeries, the behavior we obtain for
strong disorder appears consistent with recent experiments by DiCarlo
{\it et al.}, who found a nearly density independent $F$ for several
graphene samples.\cite{dicarlo07} However, this behavior departs from
the results of other numerical simulations, where a saturation value
near 0.3 was found.\cite{San-Jose07} Notice that the dip (square
samples) and the peak (rectangular samples) near the Dirac point
disappear rapidly as the disorder strength is increased and therefore
can be interpreted as signature of very weak disorder. Danneau {\it et
al.}  have recently reported the observation of ballistic behavior in
their measurements of the Fano factor for samples where $W/L \gg
1$,\cite{danneau07} although the width of the peak in their experiment
is nearly two orders of magnitude larger than the theoretical
prediction for clean samples.\cite{Tworzydlo06}

In summary, we have performed numerical simulations to evaluate the
conductivity and the shot noise Fano factor in graphene samples as a
function carrier density and in the presence of long-range
disorder. We found that the conductivity follows universal scaling
curves that depend solely on the disorder strength and the ratio of
sheet length to disorder correlation length. The Fano factor was found
to have a marked dependence on the density only for very weak disorder
and near the Dirac point. For moderate to strong disorder, the Fano
factor is nearly constant and its value is not universal. While our
model of long-range disorder applies mainly to situations where the
strong modulation of the potential in the graphene sheet is due to
roughness in the substrate, it could also be extended to account for
the effect of charge impurities present in the substrate or on the
graphene sheet. The only necessary modification is the inclusion of a
density-dependent correlation length $\xi$ to take into account
screening by carriers in the graphene sheet. However, this modulation
of $\xi$ should be weak for the low densities we have been able to
study.

We thank H. Baranger, L. DiCarlo, A. Geim, F. Guinea, P. Kim,
C. Marcus, A. Mirlin, J. R. Williams, and A. Yacoby for illuminating
discussions. C.H.L. thanks CAPES (Brazil) and the Institute of Quantum
Science and Engineering at Harvard for support.



\begin{thebibliography}{99}

\bibitem{geim_review} A. K. Geim and K. S. Novoselov, Nat. Mat.
{\bf 6}, 183 (2007).

\bibitem{RMPCastroNeto} A. H. Castro Neto {\it et al.},
arXiv:0709.1163.

%

\bibitem{Ludwig94}
A. W. W. Ludwig {\it et al.}, 
Phys. Rev. B {\bf 50}, 7526 (1994).

\bibitem{Altland06} A. Altland, Phys. Rev. Lett. {\bf 97}, 236802
(2006).

\bibitem{NomuraMacDonald07} K. Nomura and A. H. MacDonald,
Phys. Rev. Lett. {\bf 98}, 076602 (2007).

\bibitem{Ostrovsky07} P. M. Ostrovsky,
I. V. Gornyi, and A. D. Mirlin,
Phys. Rev. Lett. {\bf 98}, 256801 (2007).

\bibitem{Nomura07} K. Nomura,
M. Koshino, and S. Ryu,
Phys. Rev. Lett. {\bf 99}, 146806 (2007).

\bibitem{Bardarson07}
J. H. Bardarson {\it et al.}, 
Phys. Rev. Lett. {\bf 99}, 106801 (2007).

\bibitem{San-Jose07} P. San-Jose,
E. Prada, and D. S. Golubev, Phys. Rev. B {\bf 76}, 195445 (2007); E.
Louis {\it et al.},
{\it ibid}. {\bf 75}, 085440 (2007).

\bibitem{Titov07}
M. Titov, Europhys. Lett. {\bf 79}, 17004 (2007).

\bibitem{ryu07} S. Ryu, {\it et al.},
Phys. Rev. Lett. {\bf 99}, 116601 (2007).

\bibitem{Fradkin}
E. Fradkin, Phys. Rev. B {\bf 33}, 3257 (1986).

\bibitem{falko} V. V. Cheianov {\it et al.}, 
arXiv:0706.2968.

\bibitem{experiments}
S. Cho and M. S. Fuhrer, arXiv:0705.3239;
Y.-W. Tan {\it et al.}, 
arXiv:0707.1889; 
J. G. Checkelsky,
L. Li, and N. P. Ong, 
arXiv:0708.1959;
S. V. Morozov {\it et al.},
arXiv:0710.5304.  

\bibitem{LeeFisher81} P. A. Lee and D. S. Fisher,
Phys. Rev. Lett. {\bf 47}, 882 (1981).

\bibitem{ShonAndo98} N. H. Shon and T. Ando, J. Phys. Soc. Jap. {\bf
67}, 2421 (1998); P. M. Ostrovsky,
I. V. Gornyi, and A. D. Mirlin,
Phys. Rev. B {\bf 74} 235443 (2006).

\bibitem{Rycerz06} A. Rycerz,
J. Tworzyd{\l}o, and C. W. J. Beenakker,
Europhys. Lett. {\bf 79}, 57003 (2007).

\bibitem{schomerus07} H. Schomerus, Phys. Rev. B {\bf 76}, 045433
  (2007).

\bibitem{BreyFertig06} L. Brey and H. A. Fertig, Phys. Rev. B {\bf
73}, 235411 (2006).

\bibitem{yacoby} J. Martin {\it et al.}, 
arXiv:0705.2180;
M. Ishigami {\it et al.}, 
Nano Letters {\bf 7}, 1643 (2007).

\bibitem{Tworzydlo06}
J. Tworzyd{\l}o {\it et al.}, 
Phys. Rev. Lett. {\bf 96}, 246802 (2006).

\bibitem{blanter} Ya. M. Blanter and M. B\"uttiker, Phys. Rep. {\bf
336}, 1 (2000).

\bibitem{dicarlo07} L. DiCarlo {\it et al.}, 
arXiv:0711.3206.

\bibitem{danneau07} R. Danneau {\it et al.}, 
arXiv:0711.4306.


\end{thebibliography}
\end{document}